\documentclass[aps,showpacs,twocolumn]{revtex4-1}
\usepackage{epsfig}
\usepackage{graphicx,color,dcolumn}
\usepackage{epstopdf}
\usepackage{amsmath,amssymb}
\usepackage{multirow}
\usepackage{diagbox}
\usepackage{changes}
\usepackage{booktabs}
\usepackage{threeparttable}
\usepackage{subfigure}
\usepackage{hyperref}

\begin{document}
\title{ Dynamical study of $D^{*}DK$ and $D^{*}D \bar{D}$ systems at quark level}
\author{Yue Tan}
\email[E-mail: ]{tanyue@ycit.edu.cn}
\affiliation{Department of Physics, Yancheng Institute of Technology, Yancheng 224000, P. R. China}

\author{Xuejie Liu}
\email[E-mail: ]{1830592517@qq.com}
\affiliation{School of Physics, Henan Normal University, Xinxiang 453007, P. R. of China}

\author{Xiaoyun Chen}
\email[E-mail: ]{xychen@jit.edu.cn}
\affiliation{College of Science, Jinling Institute of Technology, Nanjing 211169, P. R.China}

\author{Youchang Yang}
\email[E-mail: ]{yangyc@gues.edu.cn }
\affiliation{Department of Physics, Guizhou University of Engineering Science, Bijie 551700, China.}

\author{Hongxia Huang}
\email[E-mail: ]{hxhuang@njnu.edu.cn (Corresponding author) }
\affiliation{Department of Physics, Nanjing Normal University, Nanjing 210023, P.R. China}
\author{Jialun Ping}
\email[E-mail: ]{jlping@njnu.edu.cn (Corresponding author)}
\affiliation{Department of Physics, Nanjing Normal University, Nanjing 210023, P.R. China}
\date{\today}

\begin{abstract}
 Inspired by that Belle\uppercase\expandafter{\romannumeral2} Collaboration recently reported  $T_{cc}$,  which can be interpreted as a molecular $DD^{*}$, we investigated the trihadron system of $T_{cc}$ partner with $IJ^{P}$=$01^{-}$ in the framework of a chiral quark model. It's widely accepted that the main component of $X(3872)$ contains the molecular $\bar{D}D^{*}$, while the main component of $D_{s0}^{*}(2317)$ is molecular $DK$. Based on these three well-known exotic states, $T_{cc} (DD^{*})$, $X(3872) (\bar{D}D^{*})$ and $D_{s0}^{*}(2317) (DK)$, we dynamically investigate  $D^{*}DK$ and $DD^{*}\bar{D}$ systems at quark level to search for possible bound states.  The results show that both of them are bound states, in which the binding energy of the molecular state $DD^*K$ is relatively small, only 0.8 MeV, while the binding energy of $DD^*\bar{D}$ is up to 1.9 MeV.  According to the calculation results of the Root-square-mean distances, the spatial structure of the two systems shows obvious ($DD^*$)-($\bar{D}$/$K$) structure, in which $D$ is close to $D^*$ while  $DD^*$  as a whole is relatively distant from the third hadron ($\bar{D}$/$K$), which are similar to the nucleon-electron structure. As a result, we strongly recommend that these bound states  $DD^*\bar{D}$ and $DD^*K$ are searched for experimentally.
\end{abstract}

\maketitle

\section{Introduction} \label{introduction}
The quark model has been an important tool for understanding the low energy region of QCD, thus unveiling the internal quark structure of hadrons is very important. At first, due to the limitation of experimental precision, the hadron states found in experiments are mainly ground states. In this case, the hadrons of these ground states can be well explained by the $q\bar{q}$ $ (q=u,d,s,c,b) $ structure. In 2003, Belle \cite{Choi:2003ue} reported the famous $X(3872)$, which opened a new era in the quark model for the reason that the  $X(3872)$ can be interpreted as a molecular $D\bar{D}^{*}$. After that, excited mesons including both radial and angular excited mesons can be explained by a $q\bar{q}^{\prime}$+$n\bar{n}$ $(n=u,d,s)$ configuration, which are also named as hidden charm or bottom tetraquark. Until the recent discovery of $T_{cc}$, the view has been changed to that multiquark system can be composed of open charm or bottom quarks, because the internal quark configuration of $T_{cc}$ is $c\bar{q}c\bar{q}$ configuration different from previous view. Therefore, it is reasonable to  believe that with the continuous improvement of experimental accuracy in the future, the exotic statse with open flavor three-meson structure ($q\bar{q}^{\prime}$-$q\bar{q}^{\prime}$-$q\bar{q}^{\prime}$) will be observed experimentally.

In fact, a lot of work \cite{Shevchenko:2015oea,MartinezTorres:2018zbl,Wu:2019vsy,Huang:2019qmw,Pan:2022xxz,Shen:2023uus,Ortega:2024ecy,Jin:2024zyy,Wu:2022wgn,Gordillo:2024sem} has been done on the calculation of tri-hardon system. In 2015, the authors \cite{Shevchenko:2015oea} take advantage of Faddeev-type AGS equations, in which two phenomenological and one chirally motivated $\bar{K}N$ potentials are used,  to perform dynamically exact calculations of a quasi-bound state in the $\bar{K}\bar{K}N$ trihadron system. The result shows that $\bar{K}N$ is good candidate of $\Lambda(1405)$, and  $\bar{K}\bar{K}N$ is good candidate of $\Xi(1950)$ state. Based on the fact that meson $D$ and $K$ can form $D_{s0}^*(2317)$, in Ref. \cite{MartinezTorres:2018zbl}, the authors search for bound states in the $DDK$ system. By coupling effects, they found a bound state with  binding energy of $DDK$ system  around $60$-$70$ MeV. According to this result, the Ref. \cite{Wu:2019vsy} applies the Gaussian expansion method to study the $DDDK$ system and show that it binds as well. Through the study of partial decay of $DDK$, the authors in Ref.\cite{Huang:2019qmw} also believe that $DDK$ is a stable state. In the framework of effective Lagrangian approach, Ref. \cite{Pan:2022xxz}  claims that the molecular state $\bar{D}\bar{D}^*$ is a possible candidate for $T_{cc}$, while the molecular state $\bar{D}^{(*)}\Sigma_c$ are a candidate for $P_c$ states.  For the reason the two molecules share a meson $\bar{D}$,  the trihadron state $\bar{D}\bar{D}^{*}\Sigma_c$ is also possible stable state. The authors \cite{Shen:2023uus} further investigate  three-hadron systems  $\eta K^{*} \bar{K}^{*}$, $\pi K^{*} \bar{K}^{*}$ and $K K^{*} \bar{K}^{*}$  within the framework of fixed-center approximation, where $K^{*} \bar{K}^{*}$ is treated as the fixed-center, corresponding to the possible scalar meson $a_0(1780)$ or the tensor meson $f_2^{\prime}(1525)$. They find several resonances with a mass around 2000 MeV.  Under the assumption that the heavy partner of the $T_{cc}$ exists as a $DD^{*}$ state, P.~G.~Ortega \cite{Ortega:2024ecy} explores  Efimov effect in the $D^{*}D^{*}D^{*}$ system and find a stable bound state.

According to the above work, there may be stable bound states in the trihadron system, which characterised by that one of meson pair can form exotic states reported by experiment. However, the three hadronic bound states obtained by these works are almost all at the hadronic level. In order to search for the stable trihadron state of $T_{cc}$ partner, we will deeply study the $DD^{*}\bar{D}$ system which is  obtained by combining $DD^{*}$ in $T_{cc}$ and $D^{*}\bar{D}$ in $X(3872)$, and  $DD^{*}K$ system which is  obtained by combining $DD^{*}$ in $T_{cc}$ and $DK$ in $D_{s0}^{*}(2317)$ within the framework of quark model. In this article, $S$-$D$ coupling effect is taken into account in the calculation to replace the contribution from color structure of three-meson structure. To calculate the hexaquark properties, we use the Rayleigh-Ritz variational method with a Gaussian expansion method (GEM), which allows us to expand each relative motion in the system in terms of Gaussian basis functions.

The structure of this paper is organized as follows:  After introduction, details of ChQM and GEM are introduced in Section II. In Section III, we present the numerical results and a method of finding.  The last section is devoted to the summary.

\section{ Theoretical framework} \label{chiral quark model}

Phenomenological models remain primary tools in unveiling the nature of experimentally observed multiquark candidates. Therefore, the $D^{*}DK$ and $D^{*}D \bar{D}$ systems are thoroughly investigated by means of a chiral quark model (ChQM). Furthermore, the Gaussian expansion method (GEM), renowned for its high accuracy in computing few-body systems, is utilized to explore the bound states of the hexaquark system.

\subsection{Chiral quark model}

In the ChQM, the Hamiltonian consists of three parts: the mass term ($m_i$), the kinetic term ($\frac{\vec{p_i}}{2 m_i}$), and the potential term ($V(\textbf{r}_{ij})$), which can be written as,
\begin{eqnarray}
\nonumber
H &=& \sum_{i=1}^6m_i+ \frac{{\vec{p}_{12}}^2}{2\mu_{12}}+ \frac{{\vec{p}_{34}}^2}{2\mu_{34}}+ \frac{{\vec{p}_{56}}^2}{2\mu_{56}}+ \frac{{\vec{p}_{12,34}}^2}{2\mu_{12,34}}+ \frac{{\vec{p}_{1234,56}}^2}{2\mu_{1234,56}} \\
&+&  \sum_{i<j=1}^6 [ V_{con}(\textbf{r}_{ij})+V_{oge}(\textbf{r}_{ij})+V_{\chi,\chi=\pi,\eta, K, \sigma}(\textbf{r}_{ij}) ],
\end{eqnarray}
where $\mu_{ij}$ and $\vec{p}_{ij}$ are the reduced mass and momentum of two interacting quarks or quark-clusters, $m_i$ is the constituent mass of $i$-th quark (antiquark), whcih can be written as follows

\begin{eqnarray}
\mu_{12}&=&\frac{m_{{1}}  m_{{2}}}{m_{{1}} + m_{{2}}},\mu_{34}=\frac{m_{{3}}  m_{{4}}}{m_{{3}} + m_{{4}}},\mu_{56}=\frac{m_{{5}}  m_{{6}}}{m_{{5}} + m_{{6}}},   \nonumber\\
\mu_{12,34}&=&\frac{(m_1+m_2)(m_3+m_4)}{m_1+m_2+m_3+m_4}, \nonumber \\
\mu_{1234,56}&=&\frac{(m_1+m_2+m_3+m_4)(m_5+m_6)}{m_1+m_2+m_3+m_4+m_5+m_6}, \nonumber \\
\vec{p}_{12}&=&\frac{m_2\vec{p}_1-m_1\vec{p}_2}{m_1+m_2},\vec{p}_{34}=\frac{m_4\vec{p}_3-m_3\vec{p}_4}{m_3+m_4}, \nonumber \\
\vec{p}_{56}&=&\frac{m_6\vec{p}_5-m_5\vec{p}_6}{m_5+m_6}, \nonumber \\
\vec{p}_{12,34}&=&\frac{(m_3+m_4)\vec{p}_{12}-(m_1+m_2)\vec{p}_{34}}{m_1+m_2+m_3+m_4}, \\
\vec{p}_{1234,56}&=&\frac{(m_1+m_2+m_3+m_4)\vec{p}_{56}-(m_5+m_6)\vec{p}_{1234}}{m_1+m_2+m_3+m_4+m_5+m_6},\nonumber
\end{eqnarray}

In our investigation, we specifically consider dynamical chiral symmetry breaking ($V_{\chi,\chi=\pi,\eta, K, \sigma}(\textbf{r}_{ij})$), color confinement ($V_{con}(\textbf{r}_{ij})$), and perturbative one-gluon exchange interactions ($V_{oge}(\textbf{r}_{ij})$), which correspond to fundamental properties of QCD in its low-energy regime. Given the three-meson system ($c\bar{q}$-$c\bar{q}$-$q\bar{s}(q\bar{c})$) with negative-parity, both $S$-wave and $D$-wave motions are present within this system. As a result, the potential included in our Hamiltonian incorporates both tensor force potential ($V^{T}(\textbf{r}_{ij})$) and spin-orbit coupling potential ($V^{SO}(\textbf{r}_{ij})$).

$V_{con}(\textbf{r}_{ij})$ represents the confining potential, mirroring the essential "confinement" characteristic of QCD. Based on lattice regularized QCD findings, it has been established that multi-gluon exchanges yield an attractive potential that increases linearly with the distance between infinitely heavy quarks. Within the quark model, three distinct potential energy forms exist: square confinement \cite{Tan:2020ldi}, linear confinement \cite{Yang:2021hrb}, and exponential confinement \cite{Vijande:2004he} with a screening effect. Square confinement and linear confinement potentials share similarities, while the exponential confinement potential is notable for its  screening effect  at higher energy region.  Given our focus on the bound state problem, there exists minimal disparity among these three potential energies. Therefore, we adopt square confinement potential in this article. The $V_{con}(\textbf{r}_{ij})$ comprises both the central force $V_{con}^{C}(\textbf{r}_{ij})$ and the spin-orbit force $V_{con}^{SO}(\textbf{r}_{ij})$.

\begin{align}
\label{Vcon}
\begin{split}
 \left \{
\begin{array}{ll}
    V_{con}^C(\textbf{r}_{ij}) &= ( -a_{c} r_{ij}^{2}-\Delta) \boldsymbol{\lambda}_i^c \cdot \boldsymbol{\lambda}_j^c\\
\\
    V_{con}^{SO}(\textbf{r}_{ij}) &= -\boldsymbol{\lambda}_i^c \cdot \boldsymbol{\lambda}_j^c \frac{a_c}{4m_i^2m_j^2}[ ((m_i^2+m_j^2)(1-2a_s) \\
    &+ 4m_im_j(1-a_s))(\vec{S}_{+}\cdot \vec{L})+\\
    &((m_j^2-m_i^2)(1-2a_s))(\vec{S}_{-}\cdot \vec{L}) ]\\
\end{array}
\right.
\end{split}
\end{align}

In Eqs.~\ref{Vcon}, the parameters $a_c$, $a_s$, and $\Delta$ are defined and listed in Table \ref{modelparameters}, and $\vec{S}_{\pm}=\vec{S}_{i}\pm\vec{S}_j$. $\boldsymbol{\lambda}^{c}$ are $SU(3)$ color Gell-Mann matrices.

Dynamical chiral symmetry breaking leads to the emergence of Goldstone boson exchange interactions among constituent light quarks $u$, $d$, and $s$. The chiral part of  Hamiltonian $V_{\chi}(\textbf{r}_{ij})$  can be resumed as follows,

\begin{eqnarray}
V_{\chi}(\textbf{r}_{ij})=V_{\pi,\eta, K, \sigma}^{C}(\textbf{r}_{ij})+V_{\pi,\eta, K}^{T}(\textbf{r}_{ij}) +V_{\sigma}^{SO}(\textbf{r}_{ij}),
\end{eqnarray}
where $C$ stands for central, $T$ for tensor, and $SO$ for spin-orbit potentials. The central part presents four different contributions,

\begin{eqnarray}
V_{\chi}^{C}(\textbf{r}_{ij})=V_{\pi}^{C}(\textbf{r}_{ij})+V_{K}^{C}(\textbf{r}_{ij})+V_{\eta}^{C}(\textbf{r}_{ij})+V_{\sigma}^{C}(\textbf{r}_{ij}),
\end{eqnarray}
where  can be written as follows,

\begin{align}
\begin{split}
 \left \{
\begin{array}{ll}
V_{\pi}^C(\textbf{r}_{ij}) & =  \frac{g^2_{ch}}{4\pi} \frac{m_{\pi}^2}{\Lambda_{\pi}^2-m_{\pi}^2} \frac{\Lambda^2_{\pi}}{\Lambda^2_{\pi}-m^2_{\pi}}m_{\pi}[ Y(m_{{\pi}}r_{ij})-\frac{\Lambda_{\pi}^3}{m_{\pi}^3}\\
&Y(\Lambda_{{\pi}}r_{ij})](\vec{\sigma}_i\cdot\vec{\sigma}_j) \sum_{a=1}^{3} \boldsymbol{\lambda}_i^a \boldsymbol{\lambda}_j^a,   \\
V_{K}^C(\textbf{r}_{ij}) & =  \frac{g^2_{ch}}{4 \pi} \frac{m_{K}^2}{\Lambda_{K}^2-m_{K}^2} \frac{\Lambda^2_{K}}{\Lambda^2_{K}-m^2_{K}}m_{K}[ Y(m_{{K}}r_{ij})-\frac{\Lambda_{K}^3}{m_{K}^3}\\
&Y(\Lambda_{{K}}r_{ij})](\vec{\sigma}_i\cdot\vec{\sigma}_j) \sum_{a=4}^{7} \boldsymbol{\lambda}_i^a \boldsymbol{\lambda}_j^a,   \\
V_{\eta}^C(\textbf{r}_{ij}) & =  \frac{g^2_{ch}}{4\pi} \frac{m_{\eta}^2}{\Lambda_{\eta}^2-m_{\eta}^2} \frac{\Lambda^2_{\eta}}{\Lambda^2_{\eta}-m^2_{\eta}}m_{\eta}[ Y(m_{{\eta}}r_{ij})-\frac{\Lambda_{\eta}^3}{m_{\eta}^3}\\
&Y(\Lambda_{{\eta}}r_{ij})](\vec{\sigma}_i\cdot\vec{\sigma}_j)[\cos\theta_{P}(\boldsymbol{\lambda}_i^8 \boldsymbol{\lambda}_j^8)-\frac{2}{3}\sin\theta_{P}],   \\
V_{\sigma}^C(\textbf{r}_{ij}) & =  \frac{g^2_{ch}}{4\pi}  \frac{\Lambda^2_{\sigma}}{\Lambda^2_{\sigma}-m^2_{\sigma}}m_{\sigma}[ Y(m_{{\sigma}}r_{ij})-\frac{\Lambda_{\sigma}}{m_{\sigma}}\\
&Y(\Lambda_{{\sigma}}r_{ij})],   \\
\end{array}
\right.
\end{split}
\end{align}
where $\boldsymbol{\lambda}^a$ are $SU(3)$ flavor Gell-Mann matrices,  $m_{\pi}$ and $m_{\eta}$ are the masses of $SU(3)$ Goldstone bosons, taken to be their experimental values; $m_{\sigma}$ is determined by the relation $m_{\sigma}^{2}\approx m_{\pi}^{2}+4m_{u,d}^{2}$. $\Lambda_{\chi}$ is the cut-offs, $g^2_{ch}/4\pi$ is the Goldstone-quark coupling constant, which is determined from the $NN\pi$ coupling constant. Finally, $Y(x)$ is the standard Yukawa function defined by $Y(x)=e^{-x}/x$, $G(x)=(1+1/x)Y(x)/x$ and $H(x)=(1+3/x+3/x^2)Y(x)/x$.

There are three different contributions to the tensor potential,
\begin{eqnarray}
V_{\chi}^{T}(\textbf{r}_{ij})=V_{\pi}^{T}(\textbf{r}_{ij})+V_{K}^{T}(\textbf{r}_{ij})+V_{\eta}^{T}(\textbf{r}_{ij}),
\end{eqnarray}
being each interaction given by,
\begin{align}
\begin{split}
 \left \{
\begin{array}{ll}
V_{\pi}^T(\textbf{r}_{ij}) & =  \frac{g^2_{ch}}{4\pi} \frac{m_{\pi}^2}{\Lambda_{\pi}^2-m_{\pi}^2} \frac{\Lambda^2_{\pi}}{\Lambda^2_{\pi}-m^2_{\pi}}m_{\pi}[ H(m_{{\pi}}r_{ij})-\frac{\Lambda_{\pi}^3}{m_{\pi}^3}\\
&H(\Lambda_{{\pi}}r_{ij})]S_{ij} \sum_{a=1}^{3} \boldsymbol{\lambda}_i^a \boldsymbol{\lambda}_j^a,   \\
V_{K}^T(\textbf{r}_{ij}) & =  \frac{g^2_{ch}}{4 \pi} \frac{m_{K}^2}{\Lambda_{K}^2-m_{K}^2} \frac{\Lambda^2_{K}}{\Lambda^2_{K}-m^2_{K}}m_{K}[ H(m_{{K}}r_{ij})-\frac{\Lambda_{K}^3}{m_{K}^3}\\
&H(\Lambda_{{K}}r_{ij})]S_{ij} \sum_{a=4}^{7} \boldsymbol{\lambda}_i^a \boldsymbol{\lambda}_j^a,   \\
V_{\eta}^T(\textbf{r}_{ij}) & =  \frac{g^2_{ch}}{4\pi} \frac{m_{\eta}^2}{\Lambda_{\eta}^2-m_{\eta}^2} \frac{\Lambda^2_{\eta}}{\Lambda^2_{\eta}-m^2_{\eta}}m_{\eta}[ H(m_{{\eta}}r_{ij})-\frac{\Lambda_{\eta}^3}{m_{\eta}^3}\\
&H(\Lambda_{{\eta}}r_{ij})]S_{ij}[\cos\theta_{P}(\boldsymbol{\lambda}_i^8 \boldsymbol{\lambda}_j^8)-\frac{2}{3}\sin\theta_{P}],   \\
\end{array}
\right.
\end{split}
\end{align}
where $S_{ij}$ is  defined by $3(\vec{\sigma}_i \cdot \hat{r}_{ij} )(\vec{\sigma}_j \cdot \hat{r}_{ij} )-\vec{\sigma}_i\cdot\vec{\sigma}_j$. The spin-orbit potential solely arises from the scalar $\sigma$ exchange between light quarks.

\begin{align}
V_{\sigma}^{SO}(\textbf{r}_{ij})   &=  -\frac{g^2_{ch}}{4\pi} \frac{\Lambda^2_\sigma}{\Lambda^2_\sigma-m^2_\sigma}\frac{m_\sigma^3}{2m_im_j}
	[ G(m_{\sigma}r_{ij}) \nonumber \\
&-\frac{\Lambda_\sigma^3}{m_\sigma^3}G(\Lambda_{\sigma}r_{ij})]\vec{L}\cdot \vec{S},
\end{align}
Beyond the scale of chiral symmetry breaking, one expects that the dynamics are influenced by QCD perturbative effects. These effects mimic gluon fluctuations around the instanton vacuum and are incorporated through the $V_{OGE}(\textbf{r}_{ij})$ which contains central force $V_{oge}^{C}(r_{ij})$, spin-orbit force $V_{oge}^{SO}(r_{ij})$ and tensor force $V_{oge}^{T}(r_{ij})$. The central potential can be written as
\begin{eqnarray}
V_{oge}^C(r_{ij}) =\frac{\alpha_s}{4} \boldsymbol{\lambda}_i^c \cdot \boldsymbol{\lambda}_{j}^c \left[\frac{1}{r_{ij}}-\frac{2\pi}{3} \frac{\boldsymbol{\sigma}_i\cdot \boldsymbol{\sigma}_j}{m_im_j} \delta(r_{ij}) \right],
\end{eqnarray}
where $\boldsymbol{\sigma}$ are the $SU(2)$ Pauli matrices, $r_{0}(\mu_{ij})=\frac{r_0}{\mu_{ij}}$ and $\alpha_{s}$ is an effective scale-dependent running coupling,
\begin{equation}
 \alpha_s(\mu_{ij})=\frac{\alpha_0}{\ln\left[(\mu_{ij}^2+\mu_0^2)/\Lambda_0^2\right]}.
\end{equation}
The $\delta(r_{ij})$ function, arising as a consequence of the non-relativistic reduction of the one-gluon exchange diagram between point-like particles, has to be regularized in order to perform exact calculations. It reads
\begin{eqnarray}
\delta{(\boldsymbol{r}_{ij})}=\frac{e^{-r_{ij}/r_0(\mu_{ij})}}{4\pi r_{ij}r_0^2(\mu_{ij})}. \nonumber
\end{eqnarray}
The spin-orbit force $V_{oge}^{SO}(r_{ij})$ and tensor force $V_{oge}^{T}(r_{ij})$ can be written as
\begin{subequations}
\label{Voge}
\begin{align}
    V_{oge}^{SO}(r_{ij}) &= -\frac{1}{16} \frac{\alpha_s\boldsymbol{\lambda}_i^c \cdot \boldsymbol{\lambda}_j^c}{4m_i^2m_j^2}[\frac{1}{r_{ij}^3}-\frac{e^{-r_{ij}/r_g(\mu)}}{r_{ij}^3}(1+\frac{r_{ij}}{r_g(\mu)})] \nonumber \\
    &[ (m_i^2+m_j^2+4m_im_j)(\vec{S}_{+}\cdot \vec{L})\nonumber \\
    &+(m_j^2-m_i^2)(\vec{S}_{-}\cdot \vec{L}) ],\\ \nonumber
V_{oge}^{T}(r_{ij}) &= -\frac{1}{16} \frac{\alpha_s\boldsymbol{\lambda}_i^c \cdot \boldsymbol{\lambda}_j^c}{4m_i^2m_j^2}[\frac{1}{r_{ij}^3}-\frac{e^{-r_{ij}/r_g(\mu)}}{r_{ij}}(\frac{1}{r_{ij}^2} \nonumber \\
&+\frac{1}{3r^2_g(\mu)} +\frac{1}{r_{ij}r_g(\mu)})]S_{ij},
\end{align}
\end{subequations}

\begin{table}[t]
\begin{center}
\caption{Quark model parameters ($m_{\pi}=0.7$ $fm^{-1}$, $m_{\sigma}=3.42$ $fm^{-1}$, $m_{\eta}=2.77$ $fm^{-1}$).\label{modelparameters}}
\begin{tabular}{cccc}
\hline\hline\noalign{\smallskip}
Quark masses   &$m_u=m_d$(MeV)       &313  \\
               &$m_{c}$(MeV)         &1728 \\
\hline
Goldstone bosons
                   &$\Lambda_{\pi}(fm^{-1})$     &4.2  \\
                   &$\Lambda_{\eta}(fm^{-1})$    &5.2  \\
                   &$g_{ch}^2/(4\pi)$            &0.54 \\
                   &$\theta_p(^\circ)$           &-15  \\
\hline
Confinement        &$a_{c}$(MeV$\cdot fm^{-2}$)                 &101  \\
                   &$\Delta$(MeV)                &-78.3\\
\hline
OGE                &$\alpha_{qq}$                &0.5723\\
                   &$\alpha_{qc}$                &0.4938\\
                   &$\alpha_{qs}$                &0.5350\\
                   &$\alpha_{cc}$                &0.3753\\
                   &$\hat{r}_0$(MeV)             &28.17 \\
                   &$\hat{r}_g$(MeV)             &34.5  \\
                   &$a_{s}$                      &0.777 \\
\hline\hline
\end{tabular}
\end{center}
\end{table}
All the parameters are determined by fitting the meson spectrum, taking into account only a quark-antiquark component. They are shown in Table~\ref{modelparameters}. 
\section{The wave function}
\begin{figure}[htp]
  \setlength {\abovecaptionskip} {-0.1cm}
  \centering
  \resizebox{0.30\textwidth}{!}{\includegraphics[width=2.0cm,height=1.5cm]{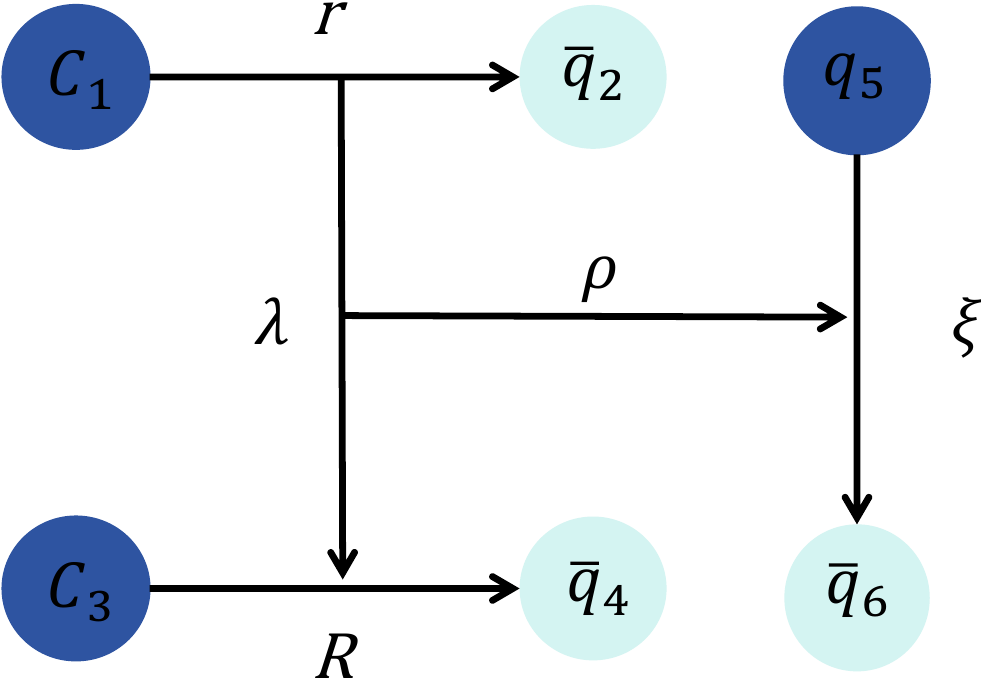}}
  \caption{The spatial configuration of $D^{*}DK$ or $D^{*}D \bar{D}$.}
\label{sptial}
\end{figure}

At the quark level, the six quarks are divided into three subclusters as shown in FIG. \ref{sptial}. In this case, $q_1$ and $\bar{q}_2$ form the first subcluster wave function $\psi(r)$; $q_3$ and $\bar{q}_4$ form the second subcluster wave function $\psi(R)$; $q_5$ and $\bar{q}_6$ form the third subcluster wave function $\psi(\xi)$; the wave function $\psi(\rho)$ represents the relative motion between the first subcluster and the second subcluster; the wave function $\psi(\lambda)$ represents the relative motion between  one cluster composed of $q_1$, $\bar{q}_2$, $q_3$ and $\bar{q}_4$  and the third subcluster. The total wave function of a hexaquark system is the internal product of color, spin, flavor and space wave functions of three sub-clusters and two spatial relative motions.

Since we are studying the hexaquark system of $T_{cc}$ partners,  two flavor wave functions are obtained,
\begin{eqnarray}
\chi^{f1} &= \frac{1}{\sqrt{2}} (c\bar{u}c\bar{d}u\bar{c} - c\bar{d}c\bar{u}u\bar{c}   ) 
\end{eqnarray}
for $D^{*}D\bar{D}$ system, 
\begin{eqnarray}
\chi^{f2} &= \frac{1}{\sqrt{2}} (c\bar{u}c\bar{d}u\bar{s} - c\bar{d}c\bar{u}u\bar{s}   ) 
\end{eqnarray}
for $D^{*}DK$ system.

As we substitute the contribution of color-excited states with $S$-$D$ coupling effects in this paper, for the total color wave function of the hexaquark of $T_{cc}$ partners, we only consider the colorless wave function obtained by three coupled color-singlet clusters, $1 \otimes 1 \otimes 1$.
\begin{eqnarray}
\chi^{c} &=& \frac{1}{\sqrt{3}} (\bar{r}r +\bar{g}g+\bar{b}b  ) \times \frac{1}{\sqrt{3}} (\bar{r}r +\bar{g}g+\bar{b}b  ) \nonumber \\
 &\times& \frac{1}{\sqrt{3}} (\bar{r}r +\bar{g}g+\bar{b}b  )
\end{eqnarray}

Since $J$ is a good quantum number, we couple the spatial orbital wave functions $\psi_{L}(\textbf{r})$ and spin wave function $\chi_S$ of the three subclusters to form $\psi_{J}(\textbf{r})$, the specific details can be found in our previous work \cite{Tan:2024pqs}. The $\psi_{J}(\textbf{r})$ of these three subclusters can be expressed as follows.
\begin{eqnarray} \label{J1J2J3}
\psi_{J_1,mJ_1}({r})= \phi_{L_1,mL_1}({r}) \otimes \chi_{S_1,mS_1}, \\
\psi_{J_2,mJ_2}({R})   = \phi_{L_2,mL_2}({R}) \otimes \chi_{S_2,mS_2}, \\
\psi_{J_3,mJ_3}({\xi})   = \phi_{L_3,mL_3}({\xi}) \otimes \chi_{S_3,mS_3}.
\end{eqnarray}
The wave function $\psi_{J_1,mJ_1}({r})$ of the first subcluster and $\psi_{J_2,mJ_2}({R})$ of the second subcluster are coupled to form the wave function $\psi_{J_{12}}$. Subsequently, this coupled wave function is coupled with the wave function $\psi_{J_3,mJ_3}({\xi})$ of the third subcluster to obtain the wave function $\psi_{J_{123}}$. Two relative motion wave functions $\phi_{L_4,mL_4}(\lambda)$ and $\phi_{L_5,mL_5}(\rho)$ are sequentially coupled with the wave function $\psi_{J_{123}}$ to finally  obtain the total orbit-spin wave function $\psi^{SO}_{i}(\textbf{r})$.

\begin{eqnarray} \label{J1J2J3}
&\psi_{J_{12},mJ_{12}}= \psi_{J_1,mJ_1}(\boldsymbol{r}) \otimes \psi_{J_2,mJ_2}(\boldsymbol{R}), \\
&\psi_{J_{123},mJ_{123}}= \psi_{J_{12},mJ_{12}}  \otimes \psi_{J_3,mJ_3}(\boldsymbol{\xi}), \\
&\psi_{JL_4,mJL_4}= \psi_{J_{123},mJ_{123}}  \otimes \psi_{L_4,mL_4}(\boldsymbol{\lambda}), \\
&\psi^{SO}_{i}(\textbf{r} )= \psi_{JL_4,mJL_4}\otimes \phi_{L_5,mL_5}(\boldsymbol{\rho}),\\ \nonumber
&i \equiv \{L_1,S_1,J_1,L_2,S_2,J_2,L_3,S_3,J_3,J_{12},J_{123},JL_4,L_4,L_5 \}.
\end{eqnarray}

During this coupling process, a total of 14 variables are involved, denoted simply as combinations represented by the index $i$. Given that our calculation focuses on the bound states of the hexaquark system of $T_{cc}$ partners, all three subclusters are in the ground state. Additionally, $J_{12}$, $J_{123}$, and $JL_4$ are all set to 1. Their intercluster motion can be either $S$-wave or $D$-wave. We have listed their possible combinations in Table \ref{JJ}. Because of no difference between spin of quark and antiquark, the meson-meson structure has the same spin wave function as the diquark-antidiquark structure. The spin wave functions of the sub-cluster are shown below,
\begin{align*}
&\chi^{11}_{\sigma}=\alpha\alpha,~~
\chi^{10}_{\sigma}=\frac{1}{\sqrt{2}}(\alpha\beta+\beta\alpha),~~
\chi^{1-1}_{\sigma}=\beta\beta,\nonumber \\
&\chi^{00}_{\sigma}=\frac{1}{\sqrt{2}}(\alpha\beta-\beta\alpha).
\end{align*}
In the context of quark spin, $\alpha$ and $\beta$ represent the third component of quark spin, taking values of $\frac{1}{2}$ and $-\frac{1}{2}$, respectively, in two distinct cases.

\begin{table}[]
\caption{ \label{JJ}  Different combinations of J-J coupling.}
\begin{tabular}{cccccccc}
\hline\hline\noalign{\smallskip}
\multicolumn{2}{c}{$J_{1}=0$} & \multicolumn{2}{c}{$J_{2}=1$} &\multicolumn{2}{c}{$J_{3}=0$}& &i   \\
      $L_1=0$ & $S_1=0$       & $L_2=0$ & $S_2=1$             & $L_3=0$ & $S_3=0$           & $L_4$=0&1       \\
      $L_1=0$ & $S_1=0$       & $L_2=0$ & $S_2=1$             & $L_3=0$ & $S_3=1$           & $L_4$=2&2       \\
\multicolumn{2}{c}{$J_{1}=1$} & \multicolumn{2}{c}{$J_{2}=1$} &\multicolumn{2}{c}{$J_{3}=0$}& &i   \\
      $L_1=0$ & $S_1=1$       & $L_2=0$ & $S_2=1$             & $L_3=0$ & $S_3=0$           & $L_4$=0&3       \\
      $L_1=0$ & $S_1=1$       & $L_2=0$ & $S_2=1$             & $L_3=0$ & $S_3=1$           & $L_4$=2&4       \\
\hline
\end{tabular}
\end{table}

Finally, the spatial wave functions are Gaussians with range parameters chosen to lie in a geometrical progression:
 \begin{align}
\phi_{nLm}(\boldsymbol{r})&=N_{nL}r^{L}e^{-(r/r_{n})^2}Y_{Lm}(\hat{r})
 \end{align}
 where $N_{nL}$ are normalization constants, which can be expressed by a general formula
\begin{align}
\label{Nnl}
N_{nL}=\left[ \frac{2^{L+2}(2\nu_n)^{L+\frac{3}{2}}}{\sqrt{\pi}(2L+1)!!} \right]^{\frac{1}{2}}.
\end{align}

\section{Results and discussions}

In this work, we primarily investigate the possibility of bound states $D^{*}D\bar{D}$ and $D^{*}DK$. Because we are more concerned about the binding energy magnitude and also to ensure comparison with experimental results, we apply a mass correction to the theoretical computed results, where the final energy is the experimental threshold plus the bounding energy ($E=Threshold_{EXP.}+B.E.$).

\begin{table}[th]
\centering
\caption{Results of calculations for the $D^{*}DK$ and $D^{*}D^{*}\bar{D}$ systems, where $c.c.1$ and $B.E.1$ represent the coupling energies and corresponding binding energy magnitudes of the two channels for the $S$-wave, respectively. Similarly, $c.c.2$ and $B.E.2$ represent the coupling energies and corresponding binding energy magnitudes of the all channels, respectively. (unit: MeV)\label{TccK}}
\begin{tabular}{cccccccccc}
\hline \hline
Index&Channel&$E$&Channel&$E$\\
\hline
1 & $DD^{*}K$                    & 4365.2  & $DD^{*}\bar{D}$                    & 5736.4  \\
2 & $D^{*}D^{*}K$                & 4507.2  & $D^{*}D^{*}\bar{D}$                & 5878.4  \\
3 & $D$-wave-$DD^{*}K$           & 4368.3  & $D$-wave-$DD^{*}\bar{D}$           & 5738.5  \\
4 & $D$-wave-$D^{*}D^{*}K$       & 4509.5  & $D$-wave-$D^{*}D^{*}\bar{D}$       & 5880.6  \\
$c.c.1$ &                          & 4364.6  &                                    & 5735.1  \\
$B.E.1$ &                          & 0.6     &                                    & 1.3     \\
$c.c.2$ &                          & 4364.4  &                                    & 5734.5  \\
$B.E.2$ &                          & 0.8     &                                    & 1.9     \\
\hline
\end{tabular}
\end{table}

\begin{table*}[htb]
\caption{\label{rms} The root-mean-square distances (unit: fm) and  contributions of all potentials to the binding energy (unit: MeV) in $D^{*}DK$ and $D^{*}D^{*}\bar{D}$ systems.}
\begin{tabular}{cccccccccccccccccccccccccccc}\hline\hline
                &knetic&con.&cl.&c.m.&$\pi$&$\eta$& $\sigma$ &&$r_{12}$ &$r_{13}$ &$r_{14}$ &$r_{15}$&$r_{16}$ &$r_{23}$&$r_{24}$&$r_{25}$&$r_{26}$&$r_{34}$&$r_{35}$&$r_{36}$&$r_{45}$&$r_{46}$&$r_{56}$ \\ \hline
$D^{*}DK$       &32.0&-1.8&-2.1&-9.1&-12.6&0.1&-7.2& &1.7    &2.3    &1.7    &5.5    &5.5    &1.7    &2.4    &5.5    &5.5    &1.7    &5.5    &5.5    &5.5    &5.5    &0.6        \\
$D^{*}D\bar{D}$ &30.0&-1.6&-1.8&-7.9&-12.8&0.5&-8.0& &1.7    &2.3    &1.7    &4.9    &4.8    &1.7    &2.4    &4.9    &4.9    &1.7    &4.9    &4.8    &4.9    &4.9    &0.6         \\
\hline\hline
\end{tabular}
\end{table*}

\begin{figure}[htb]
  \setlength {\abovecaptionskip} {-0.1cm}
  \centering
  \resizebox{0.30\textwidth}{!}{\includegraphics[width=2.0cm,height=1.5cm]{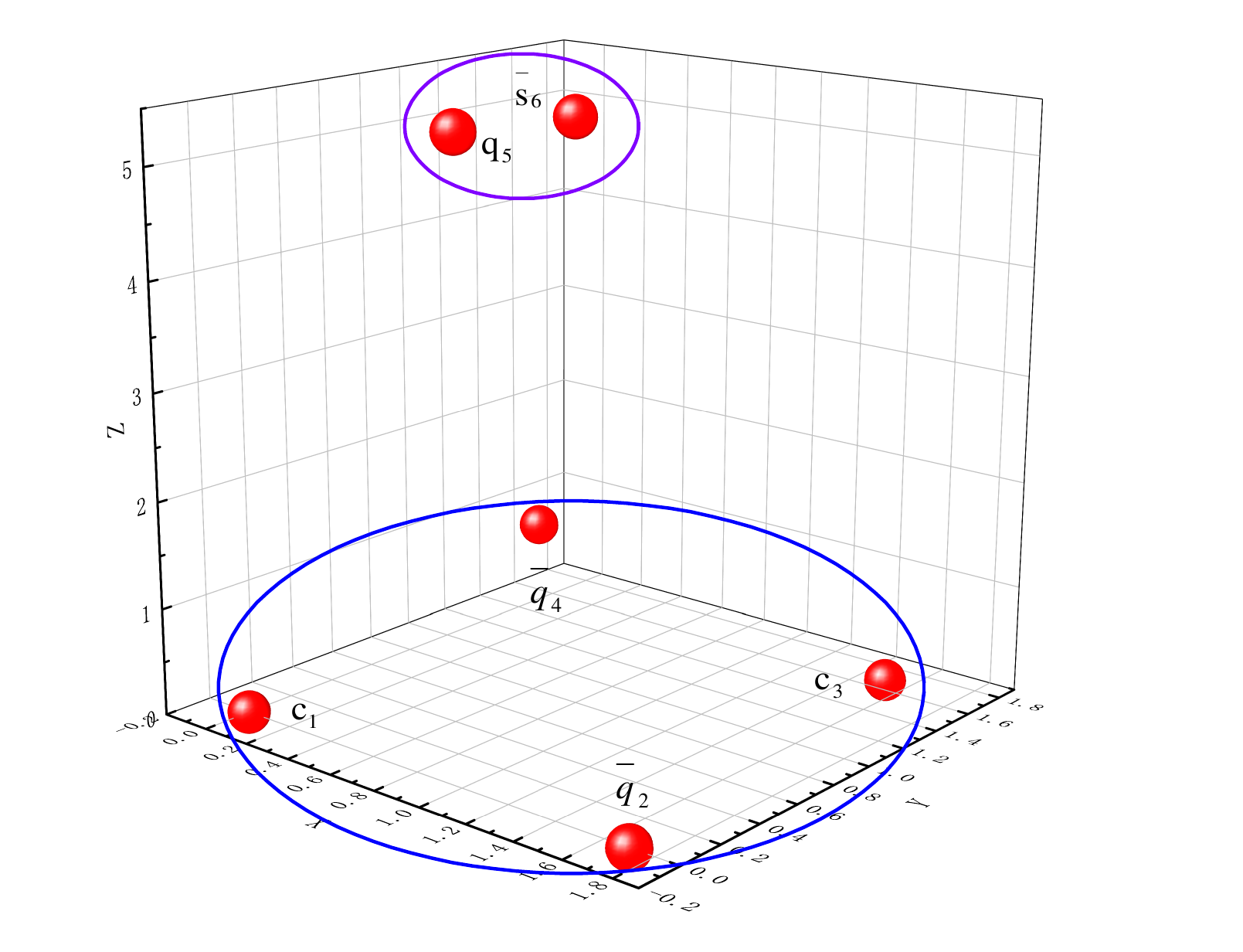}}
  \caption{The spatial structure of $D^{*}DK$.}
\label{TK}
\end{figure}

\textbf{ $D^{*}DK$ sector:} Taking into account  $IJ^{P}=01^{-}$, in the hexaquark system $c\bar{q}c\bar{q}q\bar{s}$, we have considered four physical channels: $D^{*}DK$, $D^{*}D^{*}K$, as well as $D$-wave-$D^{*}DK$ and $D$-wave-$D^{*}D^{*}K$. The results of our calculations for these channels are presented in Table \ref{TccK}. The energies of these four channels are around 4.4 GeV, and none of them are bound states. However, when we consider the coupling effects of the $S$-wave $D^{*}DK$ and $D^{*}D^{*}K$ channels, the coupling effect lowers the energy of the $D^{*}DK$ channel, which was relatively lower, below the threshold line by 0.6 MeV. Subsequently, we additionally consider the coupling effect of $S$-$D$, and the results show that the binding energy further increases to 0.8 MeV. In order to understand the binding mechanism of $D^{*}DK$, we calculated the contributions of each term in the Hamiltonian to the binding energy (the contribution of each potential energy term of the coupling energy minus the contribution of each potential energy term of the threshold channels) as well as the root-mean-square distance. The results are presented in Table \ref{rms}. The results indicate that the kinetic energy is repulsive, while most of the other potential energy terms are attractive, with the color-magnetic term, $\pi$-meson exchange, and $\sigma$-meson exchange playing a major role. This is attributed to the relatively large spacing between quarks, resulting in long-range attraction dominating the potential energy. Based on the RMS calculation, we further plotted the spatial structure of the bound state $D^{*}DK$, as shown in FIG. \ref{TK}. We can see that the first two mesons ($D^{*}D$) are relatively compact, while the third meson $K$ is slightly farther away from the first two. The overall structure resembles that of a nucleon-electron system.

\begin{figure}[htb]
  \setlength {\abovecaptionskip} {-0.1cm}
  \centering
  \resizebox{0.30\textwidth}{!}{\includegraphics[width=2.0cm,height=1.5cm]{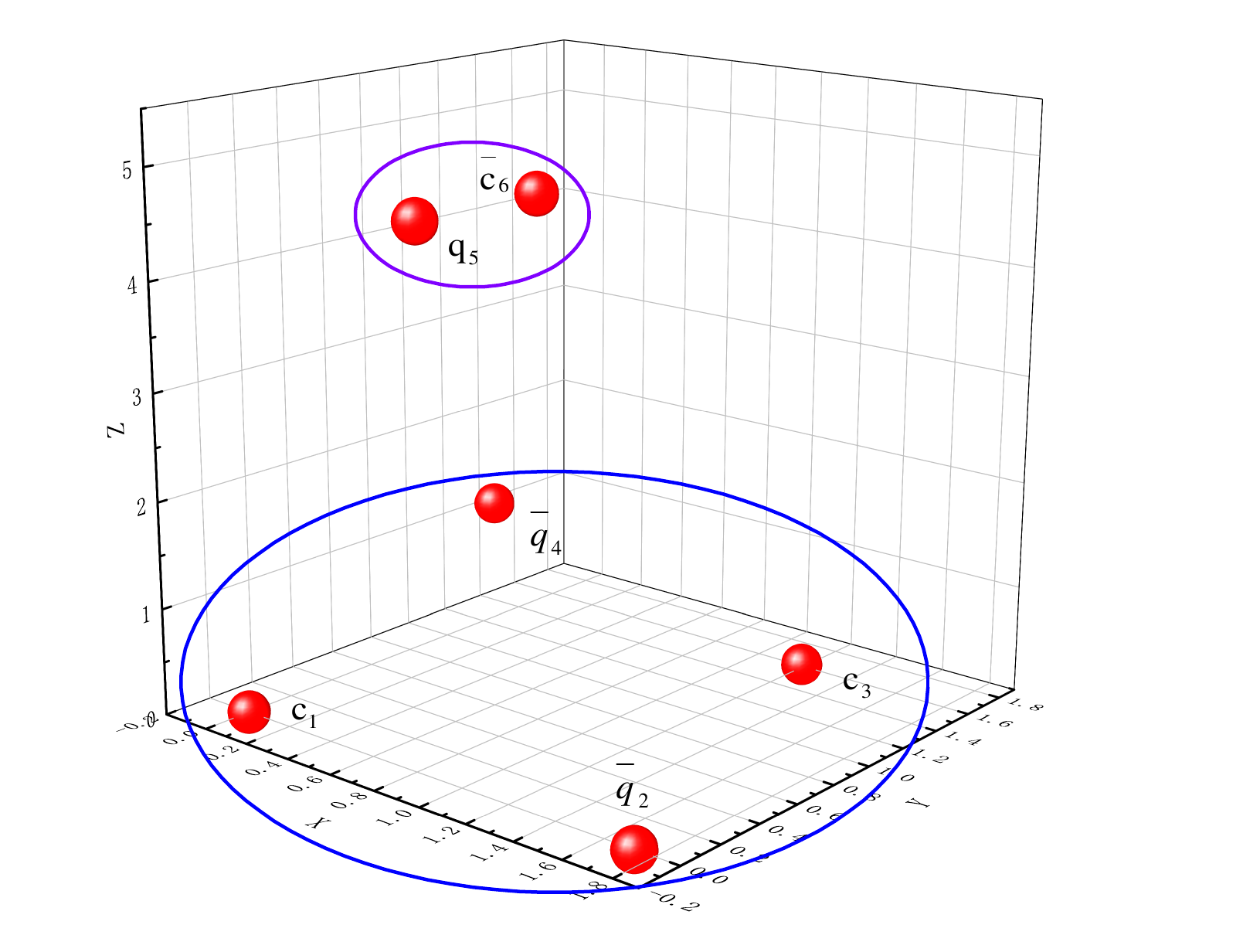}}
  \caption{The spatial structure of $D^{*}D\bar{D}$.}
\label{TD}
\end{figure}

\textbf{ $D^{*}D\bar{D}$ sector:} Similar to the $c\bar{q}c\bar{q}q\bar{s}$ system, in the $c\bar{q}c\bar{q}q\bar{c}$ system, we considered four physical channels: $D^{*}D\bar{D}$, $D^{*}D^{*}\bar{D}$, as well as $D$-wave-$D^{*}D\bar{D}$ and $D$-wave-$D^{*}D^{*}\bar{D}$. The bound state calculations show that these four single channels are all scattering states. The coupling energy of the two physical channels $D^{*}D\bar{D}$ and $D^{*}D^{*}\bar{D}$, both in the $S$-wave, is 5735.1 MeV, and their binding energy is 1.3 MeV. The binding energy obtained by the coupling of these two channels is already deeper than the binding energy of the four channels in the $c\bar{q}c\bar{q}q\bar{s}$ system. Then, we considered the contribution of the $D$-wave to the final binding energy. The calculation results show that the coupling effect of the $D$-wave increases the binding energy by an additional 0.6 MeV, resulting in a total binding energy of 1.9 MeV. Compared to the $c\bar{q}c\bar{q}q\bar{s}$ system, the quarks in the third cluster of the $c\bar{q}c\bar{q}q\bar{c}$ system are heavier, so although their kinetic energy is also repulsive, it is slightly lower than that of the former. This conclusion is supported by the root-mean-square distance of $D^{*}D^{*}\bar{D}$ in Table \ref{rms}, where the third subcluster  is closer to the first two subclusters, as shown in FIG. \ref{TD}.	

\section{Summary}

Inspired by the discovery of $T_{cc}$ ($D^{*}D$) experimentally, we combined two other well-known states, $X(3872)$ ($D^{*}\bar{D}$) and $D_{s0}^{*}$ ($DK$), to construct two hexaquark systems, $D^{*}DK$ and $D^{*}D\bar{D}$. At the quark level, utilizing the Gaussian expansion method, we performed dynamical studies on these two systems.

The calculated results reveal that both systems exhibit bound states, with the binding energy of $D^{*}DK$ slightly smaller than that of $D^{*}D\bar{D}$. Due to the similarity in quark composition between these two systems, the contributions of various terms in the Hamiltonian to their binding energies are quite similar, primarily dominated by long-range potential energy attraction and kinetic energy repulsion. The calculation of the root-mean-square distance allowed us to depict the spatial structures of these two systems, resembling that of a nucleon-electron system. As the third meson $D$ in $D^{*}D\bar{D}$ is significantly heavier than the third meson $K$ in $D^{*}DK$, the spatial structure of the $D^{*}D\bar{D}$ system is more compact, with slightly less kinetic energy repulsion compared to the $D^{*}DK$ system.

Considering the complexity of six-quark calculations, we only considered four physical channels in each system. If more excited states, such as $P$-wave and $D$-wave, are taken into account, perhaps the binding energies of these two systems would be deeper. However, the fact that the $D^{*}DK$ and $D^{*}D\bar{D}$ systems form bound states with only two channels coupled indicates their stability, strongly suggesting further experimental exploration in the future.

\acknowledgments{ This work is supported partly by the National Natural Science Foundation of China under Grant Nos. 12205249, 12205125, 11865019, 11775118 and 11535005 by the Natural Science Foundation of Jiangsu Province (BK20221166), the Science and Technology Foundation of Bijie (Grant No.BiKeLianHe[2023]17), and the Funding for School-Level Research Projects of Yancheng Institute of Technology (No. xjr2022039).}

\end{document}